\documentclass[preprint2, trackchanges, twocolumn]{aastex701}

\usepackage{mathtools}
\usepackage{todonotes}
\usepackage{xcolor}

\def\code#1{\texttt{#1}}

\defcitealias{EHTM87Paper5}{EHTC~M87*~V}
\defcitealias{EHTC_VI_M87}{EHTC~M87*~VI}

\defcitealias{EHTSgrAPaper5}{EHTC~Sgr~A*~V}
\defcitealias{EHTC_VIII}{EHTC~M87*~VIII}

\begin{document}
\title{Sensitivities of Black Hole Images from GRMHD Simulations}

\author[orcid=0000-0002-9318-9329]{Pedro Naethe Motta}
\email[show]{pedronaethemotta@usp.br}
\affiliation{Instituto de Astronomia, Geof\'{\i}sica e Ci\^encias Atmosf\'ericas, Universidade de S\~ao Paulo, S\~ao Paulo, SP 05508-090, Brazil.}
\correspondingauthor{Pedro Naethe Motta}

\author[orcid=0000-0002-6297-6549]{M\'ario Raia Neto}
\email[]{m.raia@ufabc.edu.br}
\affiliation{Universidade Federal do ABC, Santo Andr\'e, SP 09280-560, Brazil.}
\affiliation{N\'ucleo de Astrof\'isica, Universidade da Cidade de São Paulo, S\~ao Paulo, SP 01506-000, Brazil.}

\author[orcid=0000-0002-0393-7734]{Cora Prather}
\affiliation{Black Hole Initiative at Harvard University, 20 Garden Street, Cambridge, MA 02138, USA}
\email[]{cprather@fas.harvard.edu}

\author[orcid=0000-0001-9528-1826]{Alejandro C\'ardenas-Avenda\~no}
\affiliation{Department of Physics, Wake Forest University, Winston-Salem, North Carolina 27109, USA}
\email[]{cardenas@wfu.edu}

\begin{abstract}
The advent of high-fidelity imaging of supermassive black holes calls for efficient and robust data-analysis methods. In this work, we use \code{Jipole}, a differentiable, \code{ipole}-based radiative transfer code, to enable gradient-based analyses of images generated from state-of-the-art general relativistic magnetohydrodynamic (GRMHD) simulations. We compute image sensitivities, i.e., pixel-wise derivatives of the intensity with respect to model parameters, which form the Jacobian of the forward model and define a local map from parameter space to image space. Using these sensitivities in a mock data analysis, we find that GRMHD-based images generate a structured error landscape for parameter fitting, with anisotropies and local minima, making parameter exploration nontrivial but still tractable when guided by gradient information. We characterize this landscape through the Jacobian and assess the feasibility and usefulness of gradient-based recovery under idealized, blurred, and noisy conditions. Our results show that automatic differentiation-computed image gradients can guide parameter exploration effectively even in the presence of noise. These findings establish a basis for efficient, high-precision model--data comparisons in black hole imaging and motivate the integration of these sensitivities into advanced inference frameworks.
\end{abstract}

\section{Introduction}
\setcounter{footnote}{0}

General-relativistic magnetohydrodynamic (GRMHD) simulations form complete forward models useful for studying near-horizon accretion flows~\citep{Wong:2022rqr,Dhruv_2025}. Comparisons between simulation parameter studies and very long baseline interferometry (VLBI) observations have already provided constraints on accretion-flow properties and have also served as tests of the accuracy of current simulation methods \citep{EHTSgrAPaper5}. As complete models, GRMHD simulations can be compared to any observation of a source, including resolved and spectral imaging \citep{yoon2020}, jet power \citep{Tchekhovskoy_2011,Mizuno_2022}, and time variability \citep{Georgiev_2022}.

To generate predictions of the horizon-scale radio synchrotron emission observable by the Event Horizon Telescope (EHT), GRMHD simulations are post-processed with a general-relativistic radiative transfer (GRRT) code to generate a predicted intensity distribution on the sky. This imaging step is carried out with ray-tracing codes such as \code{GRay} \citep{Chan_2013}, \code{Odyssey} \citep{Pu_2016}, \code{ipole} \citep{Monika_2018,Moscibrodzka_2023}, \code{RAPTOR} \citep{Bronzwaer_2018}, \code{Mahakala} \citep{Sharma:2023nbk}, and \code{Jipole} \citep{NaetheMotta_2025_jipole}. The computational cost of this post-processing stage becomes significant when one seeks to compare each simulation over a broad space of possible imaging parameters, such as possible observer angles and electron thermodynamics models.

In \cite{NaetheMotta_2025_jipole}, we introduced a differentiable imaging framework with \code{Jipole}\footnote{Jipole is publicly available at \url{https://github.com/pedronaethe/Jipole} and archived on Zenodo \citep{pedro_naethe_motta_2026_20183201}.}, capable of computing both a simulated image and its derivatives with respect to selected input parameters. That is, for each ray-traced pixel, \code{Jipole} computes the derivatives of the specific intensity $I_\nu$ with respect to the post-processing parameters $P_i$, denoted $dI_\nu/dP_i$. As discussed in \cite{NaetheMotta_2025_jipole}, these sensitivities can enable dramatic efficiency and capability improvements for both forward and backward modeling pipelines. 

However, differentiable imaging of GRMHD models is not necessarily guaranteed to be either useful or feasible beforehand. Its success depends on how smoothly the image varies as the input parameters are changed. GRMHD-based images are disrupted by sharp emissivity transitions, emission region cuts, and discrete structure in the underlying simulation data \citep{Chan_2015}. These features can lead to irregular changes in the image under small parameter variations and may therefore complicate both local optimization and gradient-informed inference. For this reason, before sensitivities can be used in data analysis, it is necessary to establish both that they are computed correctly and that they remain informative when imaging GRMHD simulations.

In this work, for the first time, we compute these sensitivities for black hole images obtained from GRMHD snapshots, and use them to examine the questions raised above. We focus on derivatives with respect to two post-processing parameters: the observer inclination angle $\theta_{\rm o}$ and the electron-heating parameter $R_{\rm high}$. By post-processing parameters we mean image-model parameters that do not alter the underlying GRMHD simulation itself. The two parameters chosen in this paper provide examples of different types of dependence: $\theta_{\rm o}$ determines the observer's inclination, whereas $R_{\rm high}$ modifies the estimated plasma thermodynamics through an electron-heating prescription \citep{Moscibrodzka_2016}. The same framework can be extended to other post-processing parameters, including the accretion rate, black hole mass, or a generalized electron energy distribution.

We then use these sensitivities to conduct gradient-based fitting experiments as illustrative applications, rather than as a final inference strategy for black hole imaging. Existing Bayesian analyses of EHT data are typically carried out within stochastic sampling frameworks, for example with \code{Comrade.jl} \citep{Tiede_2022}, in order to account for systematic and stochastic observational uncertainties. Our goal here is narrower: we use simple conjugate-gradient experiments to test whether the derivatives $dI/d\theta_{\rm o}$ and $dI/dR_{\rm high}$ provide useful local information for parameter recovery and to assess how they reflect the structure of the imaging error surface.These example experiments are not intended to demonstrate superiority over established sampling approaches or to introduce a new inference methodology. Rather, they are designed to illustrate the information content of the GRMHD-based sensitivities and to show that these quantities can be computed in a stable and physically meaningful way in a realistic simulation setting. The focus of this work is therefore the successful computation of image sensitivities from state-of-the-art GRMHD simulations, enabling gradient-based analysis in a regime where analytic emission models are no longer applicable. The experiments should be interpreted as a controlled demonstration of these sensitivities and as a step toward their incorporation into more complete Bayesian inference pipelines.


The rest of this paper is organized as follows. In Sec.~\ref{sec:governing_equations} we present the governing equations and numerical methods used to compute both the images and their sensitivities. In Sec.~\ref{sec:validation} we validate \code{Jipole} against \code{ipole}, describe the numerical changes adopted in the present implementation, and compare sensitivities computed with finite differences and automatic differentiation. In Sec.~\ref{sec:error_landscape} we examine the normalized mean squared error (NMSE) landscape associated with the two parameters considered here, $\theta_{\rm o}$ and $R_{\rm high}$. In Sec.~\ref{sec:MockData} we present a simple mock-data analysis using these sensitivities, first for an injected image without blurring or noise, and then after adding blurring and Gaussian noise. Finally, in Sec.~\ref{sec:Discussion} we discuss the implications of these results and outline directions for future work.

\section{Governing Equations and Numerical Methods}
\label{sec:governing_equations}

In this section, we briefly summarize the main ingredients of \code{Jipole} and refer the reader to \cite{NaetheMotta_2025_jipole} for more details. To compute the radiation intensity from an accretion disk in a given spacetime as seen by a distant observer, one simulates the ray paths by solving the following system of two first-order differential equations
\begin{gather}
     \frac{dx^\mu}{d\lambda} = k^\mu ,
    \label{eq:pos_geo}
    \\
    \frac{dk^\mu}{d\lambda} = \Gamma^\mu_{\alpha \beta} k^\alpha k^\beta.
    \label{eq:mom_geo}
\end{gather}
In this formulation, $x^\mu$ denotes the four-position and $k^\mu$ the four-momentum of the photon. The geodesic is parameterized by the affine variable $\lambda$, while $\Gamma^\mu_{\alpha\beta}$ represents the connection coefficients, and the Einstein summation convention is employed.

The path of a photon is calculated by integrating this set of coupled equations backward from the camera's location. Once the path is calculated, the intensity is integrated forward along the ray for each pixel, following the covariant form of the radiative transfer equation
\begin{equation}
    \frac{d}{d\lambda}\left(\frac{I_\nu}{\nu^3}\right) = \left( \frac{j_{\nu}}{\nu^2}\right) - (\alpha_\nu \nu) \left(\frac{I_\nu}{\nu^3} \right),
    \label{eq:rad_trans_inv}
\end{equation}
where $\nu$ represents the photon frequency, $I_\nu$ denotes the specific intensity, $j_\nu$ is the specific emissivity, and $\alpha_\nu$ represents the absorption coefficient. The terms grouped in parentheses within Eq.~\ref{eq:rad_trans_inv} represent quantities that are relativistically invariant. 

For the simulations performed in this work, we focus on the Kerr geometry and $j_{\nu}$ is modeled as synchrotron emission from a thermal distribution of electrons \citep{Leung_2011}, 
\begin{align}
    j_\nu (\nu, \theta) & =\frac{\sqrt{2} \pi e^2 n_e \nu_s}{3 c K_2(1/\Theta_{\rm e})} (X^{1/2} + 2^{11/12} X^{1/6})^2 \times \notag \\ 
    & \exp(-X^{1/3}),
    \label{eq:emissivity}
\end{align}
where $X \equiv \nu/\nu_s$, $e$ is the electron charge, $c$ is the speed of light, and $n_e$ is the electron number density, and $\Theta_e$ is the dimensionless electron temperature. The function $K_2$ denotes the modified Bessel function of the second kind of order two. Here, $\nu_s$ is the characteristic synchrotron frequency, which depends on the magnetic field strength $B$ and pitch angle $\theta$ through
\begin{equation}
\nu_s = \frac{2}{9} \left(\frac{eB}{2\pi m_e c}\right) \Theta_{\rm e}^2 \sin\theta.
\end{equation}

Electron temperatures $\Theta_e$ are assigned according to a model introduced by \cite{Moscibrodzka_2016}, which calculates a temperature ratio $R$ given by
\begin{equation}
R = \frac{T_p}{T_e} = R_{\rm high} \frac{(\beta/\beta_{\rm crit})^2}{1 + (\beta/\beta_{\rm crit})^2} + R_{\rm low} \frac{1}{1 + (\beta/\beta_{\rm crit})^2},
\label{eq:r_model}
\end{equation}
where $\beta = P_{\rm gas}/P_{\rm mag}$ is the ratio of gas and magnetic pressures ($P_{\rm mag} = b^2/2$). The model parameters $R_{\rm low}$ and $R_{\rm high}$ prescribe the temperature ratios in the strongly magnetized (jet) and weakly magnetized (disk) regions, respectively, with the parameter $\beta_{\rm crit}$ defining the transition threshold, which we have set to $\beta_{\rm crit} = 1.0$ for all the analysis performed in this work. Given a local fluid internal energy density $u$ and mass density $\rho$ from a GRMHD simulation, the temperature ratio is used to split $u$ among the ions and electrons, giving the expression
\begin{equation}
\Theta_e = \frac{u}{\rho} \frac{(m_p/m_e)(\gamma_e-1)(\gamma_p-1)}{(\gamma_e-1)R + (\gamma_p-1)},
\label{eq:theta_e}
\end{equation}
where $m_p/m_e$ is the proton-to-electron mass ratio, and $\gamma_e, \gamma_p$ are the adiabatic indices for the two species.

Since we will employ GRMHD simulations for the aforementioned integrations, and these simulations are grid-based, as the photon traverses the GRMHD simulation grid cells these fluid properties are linearly interpolated during the forward integration of Eq.~\ref{eq:rad_trans_inv}. We now describe the numerical methods used to compute the image sensitivities and solve the radiative transfer equation.

\subsection{Image Sensitivities}

In this work, we focus on the derivatives of the specific intensity with respect to the post-processing parameter set $P$, denoted as $dI_\nu / dP$. In general, the parameter set $P$ can be quite large, but we will specifically consider two: the observer's inclination angle $\theta_{\rm{o}}$ and the temperature ratio parameter $R_{\rm high}$.

Assuming that equations \ref{eq:pos_geo}, \ref{eq:mom_geo} and  \ref{eq:rad_trans_inv} are continuous and differentiable, we use the formalism described in~\cite{Willkom_2018, NaetheMotta_2025_jipole} to write the sensitivities as
\begin{gather}
    \frac{d}{d\lambda} \frac{dx^\mu}{dP} = \frac{dk^\mu}{dP} ,
    \label{eq:pos_geo_thetao}
    \\
    \frac{d}{d\lambda} \frac{dk^\mu}{dP} = \frac{d}{dP}\left( \Gamma^\mu_{\alpha \beta} k^\alpha k^\beta\right),
    \label{eq:mom_geo_thetao}
    \\
    \frac{d}{d\lambda} \frac{d}{dP}\left(\frac{I_\nu}{\nu^3}\right) = \frac{d}{dP} \left[\left( \frac{j_{\nu}}{\nu^2}\right) - (\alpha_\nu \nu) \left(\frac{I_\nu}{\nu^3} \right) \right].
    \label{eq:rad_trans_thetao}
\end{gather}

Accounting for the fact that the connection coefficients $\Gamma^\mu_{\alpha\beta}$ are functions of the position $x^\mu$, and noting that the specific emissivity $j_\nu$ and absorption $\alpha_\nu$ depend on the photon’s state space $(x^\mu, k^\mu)$ and parameters $P$, reflecting how the underlying plasma properties vary with position, we rewrite Eqs.~\eqref{eq:mom_geo_thetao} and~\eqref{eq:rad_trans_thetao} as
\begin{align}
    \frac{d}{d\lambda} \frac{dk^\mu}{dP} 
    &= \frac{d}{dP}f_1(x^\mu, k^\mu, P) \notag \\
    &= \frac{\partial f_1}{\partial x^\mu} \frac{dx^\mu}{dP}
     + \frac{\partial f_1}{\partial k^\mu} \frac{d k^\mu}{d P} + \frac{\partial f_1}{\partial P}, 
    \label{eq:mom_geo_simp_thetao} \\
    \frac{d}{d\lambda} \frac{d}{dP}\left(\hat{I}_\nu\right) 
    &= \frac{d}{dP}f_2(x^\mu, k^\mu, \hat{I}_\nu, P) \notag \\
    &= \frac{\partial f_2}{\partial x^\mu} \frac{dx^\mu}{dP}
     + \frac{\partial f_2}{\partial k^\mu} \frac{d k^\mu}{dP}
     + \frac{\partial f_2}{\partial \hat{I}_\nu} \frac{d\hat{I}_\nu}{d P} + \frac{\partial f_2}{\partial P},
    \label{eq:rad_trans_simp_thetao}
\end{align}
where the invariant intensity is defined by $\hat{I}_\nu = I_\nu/\nu^3$, and the functions $f_1$ and $f_2$ represent the right hand side of Eqs.~\eqref{eq:mom_geo} and~\eqref{eq:rad_trans_inv}, respectively. One of the main features of \code{Jipole}, is that we use automatic differentiation (AD) to compute these partial derivatives, using the \textit{ForwardDiff.jl} package~\citep{RevelsLubinPapamarkou2016}. Thus, \code{Jipole} autodifferentiates the 4-position and 4-momentum differentials ($dx^\mu/dP$ and $dk^\mu/dP$) backwards alongside the geodesic integration, while $d\hat{I}_\nu/dP$ is integrated forward alongside the radiative transfer integration. 

Another possible approach would be to treat the entire image-generation process as a differentiable function and compute sensitivities via automatic differentiation through the ODE solver. While this can work for simple systems, and although we have not investigated this approach, it should in principle produce the similar sensitivities to the explicit differential formulation. However, contributions from numerical implementation details may reduce the stability and interpretability of the total sensitivities. We therefore follow the explicit differential approach instead, following the method used in our previous work \citep{NaetheMotta_2025_jipole}. By doing so, we ensure that the image sensitivities contain only physical contributions, remaining interpretable smooth and for parameter estimation.

Having written all the main equations, we will now provide the details of their numerical solutions and outline the key technical considerations for the implementation of \texttt{Jipole}.

\subsection{Geodesics Integration}

We utilize the second-order Runge--Kutta (RK2) method to integrate the geodesic equation, while $dx^\mu/dP$, $dk^\mu/dP$ are integrated using the Euler method as
\begin{gather}
   \left( \frac{dx^\mu}{dP}\right)_{n + 1} = \left(\frac{dx^\mu}{dP}\right)_{n} - d\lambda\ \left(\frac{dk^\mu}{dP}\right)_{n},
   \label{eq:4pos-deriv-integration-method}
   \\
    \left( \frac{dk^\mu}{dP}\right)_{n+1} = \left(\frac{dk^\mu}{dP}\right)_{n} - d\lambda\ (A_1)_{n},
    \label{eq:4mom-deriv-integration-method}
\end{gather}
where $A_1$ is the right-hand side of Eq.~\eqref{eq:mom_geo_simp_thetao}. The minus signs in these expressions arise because we are integrating the geodesics backward from the camera. 

\subsection{Intensity Integration}

We integrate the radiative transfer equation (Eq.~\ref{eq:rad_trans_inv}) by adopting the arithmetic mean for the emissivity and absorption coefficients across each integration step $j_\nu^{\rm avg} = (j_\nu^{n} + j_\nu^{n+1})/2$ and $\alpha_\nu^{\rm avg} = (\alpha_\nu^{n} + \alpha_\nu^{n+1})/2$, where $n$ denotes the integration step. By treating these averaged coefficients as constant over a single step, Eq.~\ref{eq:rad_trans_inv} admits the following local analytical solution:
\begin{equation}
    \hat{I}_{n + 1} = \hat{I}_{n} e^{d\lambda \ \alpha^{\rm avg}_\nu} + \frac{j_\nu^{\rm avg}}{\alpha_\nu^{\rm avg}} (1 - e^{d\lambda \ \alpha^{\rm avg}_\nu}).
    \label{eq:Int_solve_approx}
\end{equation}
Alongside the intensity integration, we also integrate $d\hat{I}_\nu/dP$ by means of the Euler method as
\begin{equation}
        \left( \frac{d\hat{I}}{dP}\right)_{n+1} = \left(\frac{d\hat{I}}{dP}\right)_{n} + d\lambda_{\rm cgs}\ (A_2)_{n},
        \label{eq:Intensity-deriv-integration-method}
\end{equation}
where $A_2$ denotes the right-hand side of Eq.~\ref{eq:rad_trans_simp_thetao}. The step size in ${\rm cgs}$ units is $d\lambda_{\rm cgs} = d\lambda \times [L_{\rm unit} h/(m_e c^2)]$, where $L_{\rm unit}$ is the length scale dictated by the mass of the black hole, $h$ is Planck's constant in $\rm {ergs} \cdot \rm{s}$, $m_e$ is the electron mass in grams, and $c$ is the speed of light in $\rm{cm}/\rm{s}$. This scaling is done when using the step size during the intensity integration, as it is expressed in physical units. 

The boundary condition for Eqs. \eqref{eq:4pos-deriv-integration-method} and \eqref{eq:4mom-deriv-integration-method} are specified at the camera location, where ray integration is initialized. For the invariant intensity derivative (Eq. \eqref{eq:Intensity-deriv-integration-method}) we set it to zero, $d\hat{I}_\nu/dP|_{\rm obs} = 0$, since at the first point of the integration of the radiative transfer equation, the intensity will be zero regardless of $P$. The geodesic derivatives $dx^\mu/dP$ and $dk^\mu/dP$ are obtained by differentiating the full camera construction procedure. In practice, the camera position and photon four-momentum are generated using the same coordinate transformations as in the standard \code{ipole} setup, and we compute their derivatives by automatic differentiation through this pipeline. This includes the dependence of the observer position (e.g., on parameters such as the inclination $\theta_o$), the construction of the local orthonormal tetrad, the definition of pixel-dependent directions in the image plane, the normalization of the null vector, and the transformation from the tetrad to the coordinate basis.

\section{Validation of \texttt{Jipole} with GRMHD Simulations}
\label{sec:validation}

Because \code{Jipole} uses an essentially similar algorithm to \code{ipole}~\citep{Monika_2018}, augmented with automatic-differentiation capabilities for parameter sensitivities, we first verify that our forward-model images remain consistent with those produced by \code{ipole}. We further validate the automatic differentiation sensitivities, $dI/dP$, by comparing them against finite-difference estimates to confirm the correctness of the computed derivatives. The validation steps presented in this section are solely designed to ensure the reliability of the calculations presented in this work.

\subsection{Image Consistency with \code{ipole}}

Because \code{Jipole} is not a direct one-to-one reimplementation of \code{ipole}~\citep{Monika_2018}, but instead modifies parts of the numerical implementation while preserving the same underlying radiative-transfer framework, it is necessary to verify that these changes do not alter the computed observables. We will perform the validation by imaging the relativistic thermal synchrotron emission from one snapshot of a GRMHD simulation\footnote{\code{Jipole} reads HDF5 files structured identically to the outputs produced by \code{iharm3D}~\cite{Prather_2021}, which is the most common data file representation adopted by the EHTC for GRMHD pipeline analyses.}. In particular, we use the same GRMHD snapshot\footnote{The GRMHD snapshot is publicly available; see \cite{dataverse_GRMHDsnapshot}.} employed in \cite{Prather_2023}, generated with the \code{iharm3D} code \citep{Prather_2021}, modeling accretion onto a Kerr black hole with dimensionless spin $a_\ast = 0.9375$. The simulation is three-dimensional, with a resolution of $288 \times 128 \times 128 $, and is initialized from a Fishbone--Moncrief torus with a standard and normal accretion (SANE) magnetic field configuration. The snapshot corresponds to a simulation time of $4500\, r_{\rm g}/c$, when the inner accretion flow has reached a quasi-steady state. This snapshot exactly resembles the simulations in the library utilized in \citealt{EHTM87Paper5, EHTC_VIII}, referred to here as \citetalias{EHTM87Paper5, EHTC_VIII}, respectively.

The adopted parameters are consistent with $\rm{M87}^\ast$ to mirror the simulation frameworks of \citetalias{EHTM87Paper5, EHTC_VI_M87, EHTC_VIII}, being
\begin{subequations}
\begin{gather}
    M_{\rm BH} = 6.2 \times 10^9\, M_\odot, \\
    D = 16.9\, \rm{Mpc}, \\
    \nu = 230\, \rm{GHz},
\end{gather}
\end{subequations}
where $M_{\rm BH}$ is the black hole mass expressed in units of solar masses, $D$ is the distance to the source, and $\nu$ is the monochromatic observing frequency used to form the image. The camera parameters adopted are
\begin{subequations}
\begin{gather}
    \theta_{\rm{o}} = 163^\circ, \\
    \phi = 0^\circ, \\
    \mathrm{FOV} = 160\,\mu\mathrm{as}, \\
    D_x = 44.17\, r_\mathrm{g}, \\
    N_x = N_y = 320 .
\end{gather}
\end{subequations}
where $\phi$ is the azimuthal angle of the camera, $\mathrm{FOV}$ denotes the field of view in microarcseconds, and $D_x$ represents the physical width of the camera plane measured in gravitational radii, $r_\mathrm{g} = GM_{\rm BH}/c^2$, and $G$ is the gravitational constant. The number of pixels along the two dimensions of the camera plane is denoted by $N_x$ and $N_y$. 

The physical accretion rate will depend on another scale factor that will determine the density of the gas and the strength of magnetic fields in physical units. This parameter is referred to as the mass unit parameter $\mathcal{M}$, characterizing the unit-free variables of the code in physical units. For this test, we use $\mathcal{M} = 1.672\times 10^{26}\, g$ in order to generate a total compact unpolarized flux of$\approx 0.5\, \rm{Jy}$, reproducing the EHT image libraries fit of compact flux density \citep{Wong:2022rqr}. Lastly, the electron heating model follows the prescription defined in Eq.~\ref{eq:r_model}, with $R_{\rm high} = 20$, $R_{\rm low} = 1$, $\gamma_{\rm e} = 4/3$, $\gamma_{\rm{p}} = 5/3$, $\gamma = 13/9$, and $\beta_{\rm crit} = 1$, which are standard values used in the simulation library of, e.g., \citealt{EHTM87Paper5}.

\begin{figure*}
    \centering
    \includegraphics[width=\linewidth]{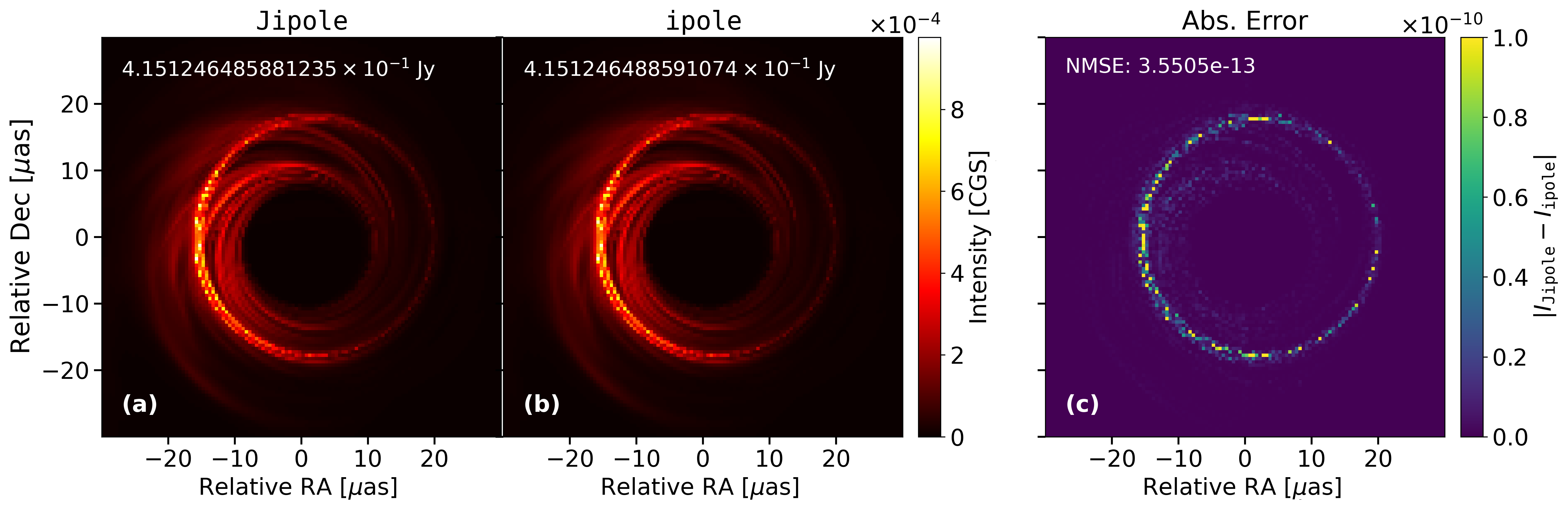}
    \caption{Image comparison between \code{Jipole} and \code{ipole} using a GRMHD snapshot. Panel (a): Intensity image computed by \code{Jipole} for the parameters specified in Section~\ref{sec:validation}. Panel (b): Intensity image computed by \code{ipole} for the parameters specified in Section~\ref{sec:validation}. For each image, the total unpolarized compact flux is shown on top of the image. Panel (c): The pixel-wise absolute difference between \code{Jipole} and \code{ipole}, defined as $\lvert I_{\rm \code{ipole}} - I_{\rm \code{Jipole}} \rvert$. The resolution of these images is $320 \times 320$.
    The camera is constructed with a field of view of $160 \mu$as; however, for visualization purposes we show a zoomed-in region corresponding to a field of view of $60\, \mu$as, since most of the larger field of view contains negligible emission}. 
\label{fig:GRMHD_comparison}
\end{figure*}

Figure \ref{fig:GRMHD_comparison} shows the results of this comparison through a $320 \times 320$ pixel image. The intensity maps produced by \code{Jipole} (Panel a) and \code{ipole} (Panel b) are visually indistinguishable, capturing the same features of the asymmetric photon ring. The total unpolarized compact flux values, printed at the top of the panels, match to seven decimal places ($F_\nu \approx 0.4151246$ Jy). The absolute difference between the two images is shown in Panel (c). The residuals are extremely small, with peak values on the order of $10^{-9}$ Jy/pixel primarily localized at the ring feature, where the integration requires many more iterations, hence amplifying errors, as discussed in~\cite{NaetheMotta_2025_jipole}. These differences are primarily driven by the use of distinct implementations of the connection coefficients, which introduce variations in the evaluation of the geodesic equations. As a result, we do not expect bit-for-bit agreement between the two codes. The numerical consistency is confirmed by a normalized mean squared error (NMSE), calculated as
\begin{equation}
    \rm{NMSE} = \frac{\sum (I_{\rm \code{Jipole}} - I_{\rm {ref}})^2}{\sum I_{\rm {ref}}^2}.
    \label{eq:NMSE}
\end{equation}
where ``ref'' represents the reference value. The NMSE is  $2.0153 \times 10^{-13}$, indicating that \code{Jipole} reproduces \code{ipole}'s behavior to better than the truncation error or other imaging uncertainties.

\subsection{Auto differentiation vs Finite Differences}

Now that we have ensured the correctness of the intensity computation against \code{ipole}, we perform the validation of the AD algorithm used to compute the sensitivities $dI/d\theta_{\rm{o}}$ and $dI/dR_{\rm high}$. To do so, we compare the computed derivatives against a finite difference (FD) approximation
\begin{equation}
    \frac{\partial I}{\partial p} = \frac{I(p + \epsilon) - I(p - \epsilon)}{2\epsilon},
\end{equation}
where $p \in P$ represents the parameter of interest, and $\epsilon = 10^{-6}$ is the small perturbation to the parameter.

We begin by evaluating $dI/dR_{\rm high}$. The R-model is used to set the plasma temperature; therefore, changes in its parameters do not affect the photon trajectories ($dx^\mu/dR_{\rm Righ} = 0$ and $dk^\mu/dR_{\rm high} = 0$). Instead, changes in $R_{\rm high}$ influence only the radiative transfer along the geodesic (Equation~\ref{eq:rad_trans_inv}). Under this scenario, Equation~\ref{eq:rad_trans_simp_thetao} simplifies and can be written as
\begin{equation}
    \frac{d}{d\lambda} \frac{d}{dR_{\rm high}}\left(\hat{I}_\nu\right) = \frac{\partial f_2}{\partial\hat{I}_\nu}\frac{d\hat{I}_\nu}{dR_{\rm high}} +\frac{df_2}{dR_{\rm high}}.
\end{equation}

\begin{figure*}
    \centering
    \includegraphics[width=\linewidth]{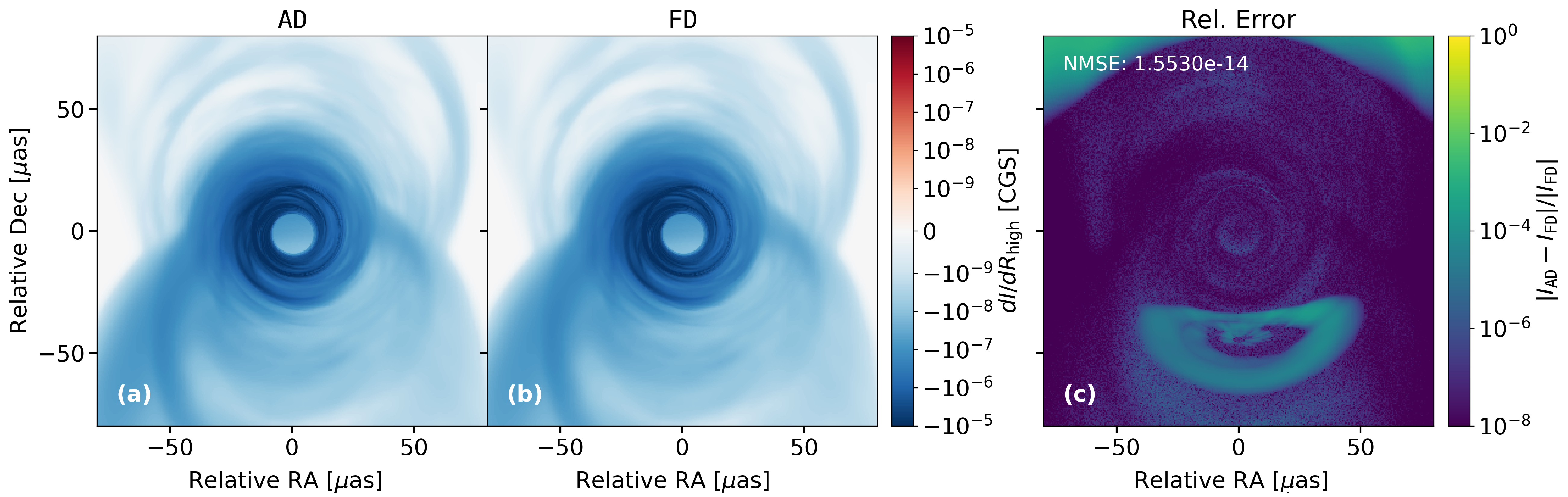}
    \caption{Comparison of the intensity derivative $dI/dR_{\rm high}$ for each pixel computed with \code{Jipole}. Panel (a) shows the result obtained using automatic differentiation (AD), while panel (b) displays the result using finite-difference (FD). The values of this sensitivities are presented on a symmetric logarithmic scale in both cases. Panel (c) illustrates the relative difference between the two methods, with the corresponding NMSE (Equation~\eqref{eq:NMSE}) reported in the top-left corner.}
\label{fig:AD_FD_Rhigh_comparison}
\end{figure*}

\begin{figure*}
    \centering
    \includegraphics[width=\linewidth]{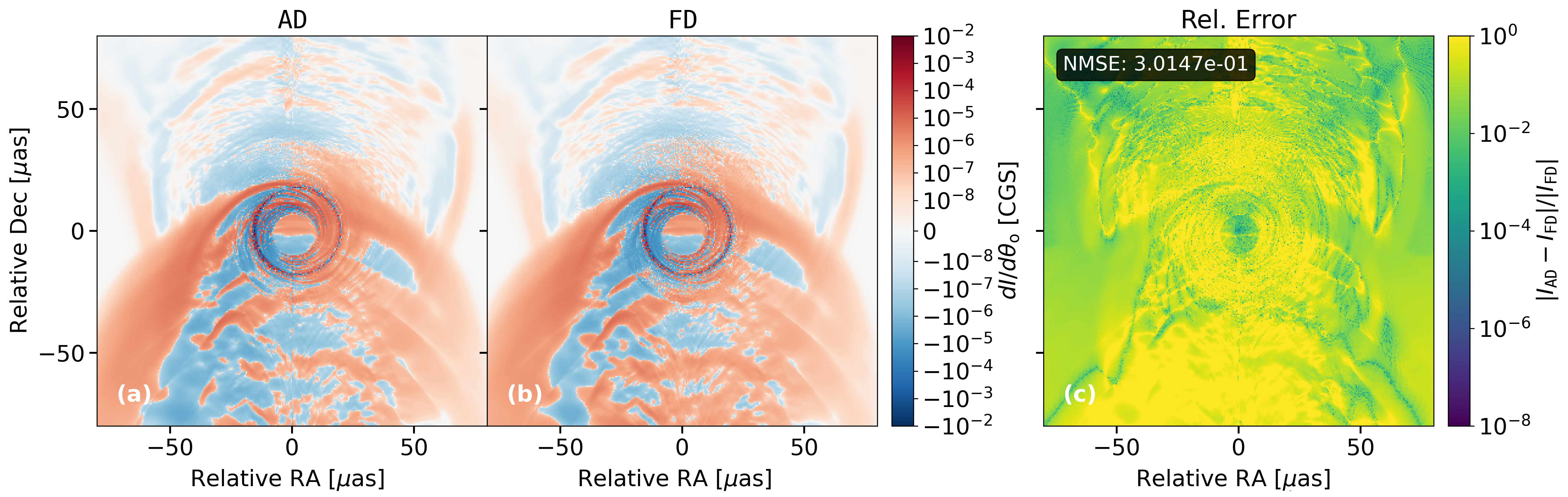}
    \caption{Comparison of the intensity derivative $dI/d\theta_{\rm{o}}$ for each pixel as computed by \code{Jipole}. Panel (a) shows the result obtained using automatic differentiation (AD), while panel (b) displays the finite-difference (FD) estimate; both are presented on a symmetric logarithmic scale. Panel (c) illustrates the logarithmic relative difference between the two methods, with the corresponding NMSE (Equation~\eqref{eq:NMSE}) reported in the top-left corner. These images have a resolution of $320\times 320$.}
\label{fig:AD_FD_th_comparison}
\end{figure*}

In Figure~\ref{fig:AD_FD_Rhigh_comparison}, we show $dI/dR_{\rm high}$ for each pixel computed via AD and FD (panels a and b), together with their relative difference (panel c), using the same GRMHD snapshot as in Section~\ref{sec:validation} at $\theta_{\rm{o}} = 163^\circ$. We observe excellent agreement between the two methods, with a NMSE of $\sim 10^{-14}$. As discussed above, variations in $R_{\rm high}$ do not affect the photon trajectories, so both AD and FD evaluate the radiative transfer along the same geodesics. The two approaches, therefore, differ only in how the derivative of the transfer equation is computed. Because the geometric part of the problem remains unchanged, numerical discrepancies are small, leading to the very small NMSE observed.

We now turn to the derivative with respect to the observer inclination, which should differ from the previous one, as photon trajectories are now modified. Figure~\ref{fig:AD_FD_th_comparison} presents $dI/d\theta_{\rm{o}}$ for each pixel computed via AD and FD (panels a and b), as well as their relative difference (panel c), using the same parameters as above. We find good overall agreement between the two methods, with a NMSE of $\sim 0.3$. In panel (c), the regions with higher relative error trace the locations where $dI/d\theta_{\rm{o}}$ changes sign, which naturally amplifies fractional differences. 

As discussed in \citet{NaetheMotta_2025_jipole}, the FD method evaluates the intensity at two nearby $\theta_{\rm{o}}$ values, which requires integrating along two distinct geodesic trajectories. This perturbation changes the null geodesics connecting the observer to the emitting plasma, causing the rays to sample different regions of the accretion flow, which will also yield different interpolated values. In contrast, AD propagates differential information along a \emph{single} geodesic trajectory. Consequently, the two methods probe slightly different physical configurations, and localized discrepancies can arise, particularly in regions where geodesics are highly sensitive to initial conditions, i.e., around the photon ring. We stress that this difference is not solely due to the finite-difference step size $\epsilon$ or machine truncation error, but also reflects the fundamentally different way in which FD and AD sample the underlying geodesic and plasma configurations for the calculations of $dI/d\theta_o$.

Lastly, in this work, we restrict the validation to Stokes $I$ only, rather than the full Stokes vector $(I, Q, U, V)$. This choice is primarily motivated by methodological simplicity, as this study represents, to our knowledge, the first application of automatic differentiation to image-space sensitivities in GRMHD radiative transfer. At this stage, our goal is to establish the stability and usefulness of the resulting gradients for parameter exploration in the simplest possible setting. In addition, extending the framework to polarized radiative transfer would substantially increase computational cost, since derivatives of all Stokes components must be propagated simultaneously. Having validated the approach in this reduced case, extending it to full polarization is a natural next step.

\subsection{The geodesic integration step size and the sigma cut-off}

We will now demonstrate how specific technical choices that are suitable for standard ray-tracing can introduce discontinuities or non-smooth behavior when computing sensitivities. In this section, we will describe their impact and the steps taken to address these issues. We will begin by discussing the integration step size used during geodesic integration, followed by the treatment of the highly magnetized jet region in GRMHD simulations.

Let us start with the integration step size used during the geodesic integration. The Illinois version of \code{ipole}~\citep{Wong:2022rqr, Prather_2023} modified the geodesic step size calculation:

\begin{subequations}
\label{eq:stepsize_ipole}
\begin{gather}
\Delta\lambda = \frac{1}{\Delta\lambda_r^{-1} + \Delta\lambda_\theta^{-1} + \Delta\lambda_\phi^{-1}}, 
\label{eq:total_stepsize}\\
\Delta\lambda_r = \frac{\epsilon_{\rm res}\,(10\,d_{\rm eh})}{|k^r|}, \\
\Delta\lambda_\theta = \frac{\epsilon_{\rm res}\, d_{\rm fac}}{|k^\theta|}, \\
\Delta\lambda_\phi = \frac{\epsilon_{\rm res}}{|k^\phi|}.
\end{gather}
\end{subequations}

In these equations, $k^\mu$ is the four-momentum of the photon and $x^\mu$ its spacetime position. The quantity $d_{\rm eh} = \min(|x^r - x^r_{\rm start}|,\,0.1)$ provides a measure of the radial distance to the event horizon and acts to progressively reduce the integration step as the photon approaches it. 

The factor $d_{\rm fac}$ controls the step size in the polar direction by measuring the proximity of the trajectory to the coordinate singularities at $\theta = 0$ and $\theta = \pi$. It is defined as a function of the normalized angular distance to the pole, 
\begin{equation}
d_{\rm pole} = \min\!\left(\frac{x^\theta}{\Delta x^\theta},\,\frac{\Delta x^\theta - x^\theta}{\Delta x^\theta}\right).
\end{equation}
which varies from zero at the poles to order unity away from them. The function $d_{\rm fac}$ is then constructed as
\begin{equation}
d_{\rm fac} =
\begin{cases}
\dfrac{d_{\rm pole}}{3}, & d_{\rm pole} < \theta_{\rm cut}, \\
\min\!\left(\dfrac{\theta_{\rm cut}}{3} + 10\,(d_{\rm pole} - \theta_{\rm cut}),\,1\right), & d_{\rm pole} \ge \theta_{\rm cut},
\end{cases}
\end{equation}
such that the step size is strongly suppressed near the poles ($d_{\rm pole} \ll 1$) through a linear scaling and then rapidly increases away from the poles, saturating at unity. The parameter $\epsilon_{\rm res}=0.01$ sets the overall scaling of the integration step.

This prescription was originally introduced for the thin-disk test conducted in \cite{Prather_2023}, where the step size is designed to decrease more aggressively toward the polar regions. This aimed to improve the accuracy of the parallel transport of the linear polarization direction along the geodesics near the poles. However, as discussed, the step size is reduced once a geodesic crosses a prescribed polar-angle threshold and increases rapidly once it crosses it outwardly. This abrupt change in the integration step tends to lead to non-smooth behavior when computing sensitivities.

To avoid this spurious behavior, we instead adopt the step size prescription used in \code{grmonty}~\citep{Dolence_2009} and \code{GPUmonty}~\citep{NaetheMotta_2026_GPUmonty}, hereafter referred to as $\Delta\lambda_{\rm new}$, in contrast to the original \code{ipole} prescription which we denote as $\Delta\lambda_{\rm ipole}$. In this approach, the step size $\Delta\lambda_{\rm new}$, is determined using Equation~\eqref{eq:total_stepsize} but taking

\begin{subequations}
\label{eq:stepsize_grmonty}
\begin{gather}
\Delta\lambda_r = \frac{\epsilon_{\rm res}}{|k^r|}, \\
\Delta\lambda_\theta = \frac{\epsilon_{\rm res} \, \min(x^\theta,\, 1 - x^\theta)}{|k^\theta|}, \\
\Delta\lambda_\phi = \frac{\epsilon_{\rm res}}{|K^\phi|}.
\end{gather}
\end{subequations}

In particular, the factor $\mathrm{min}(x^\theta, 1 - x^\theta)$ is added to reduce the step size near the polar axis in a continuous manner, without introducing non-differentiable behavior. 

Another source of spurious behavior in the sensitivities stems from the treatment of the highly magnetized jet regions. In GRMHD simulations, the polar funnel region is often poorly constrained because numerical floors are imposed on the density and internal energy to maintain stability in very low-density regions~\citep{Wong:2022rqr}. As a result, the temperatures and densities in the funnel can become artificially large compared to the expected physical values. A common practical approach in post-processing is therefore to exclude highly magnetized regions from the radiative transfer calculation by applying a magnetization threshold ($\sigma_{\rm cut}$) based on the plasma parameter $\sigma = b^2/\rho$, where $b^2$ is the magnetic strength squared and $\rho$ is the mass density of the gas. Emission is artificially set to zero where $\sigma$ exceeds a chosen cutoff value. While this prescription helps avoid nonphysical emission from the funnel region, it introduces a discontinuity in the emissivity that can affect the resulting spectra and sensitivities.

\begin{figure*}
    \centering
    \includegraphics[width=\textwidth]{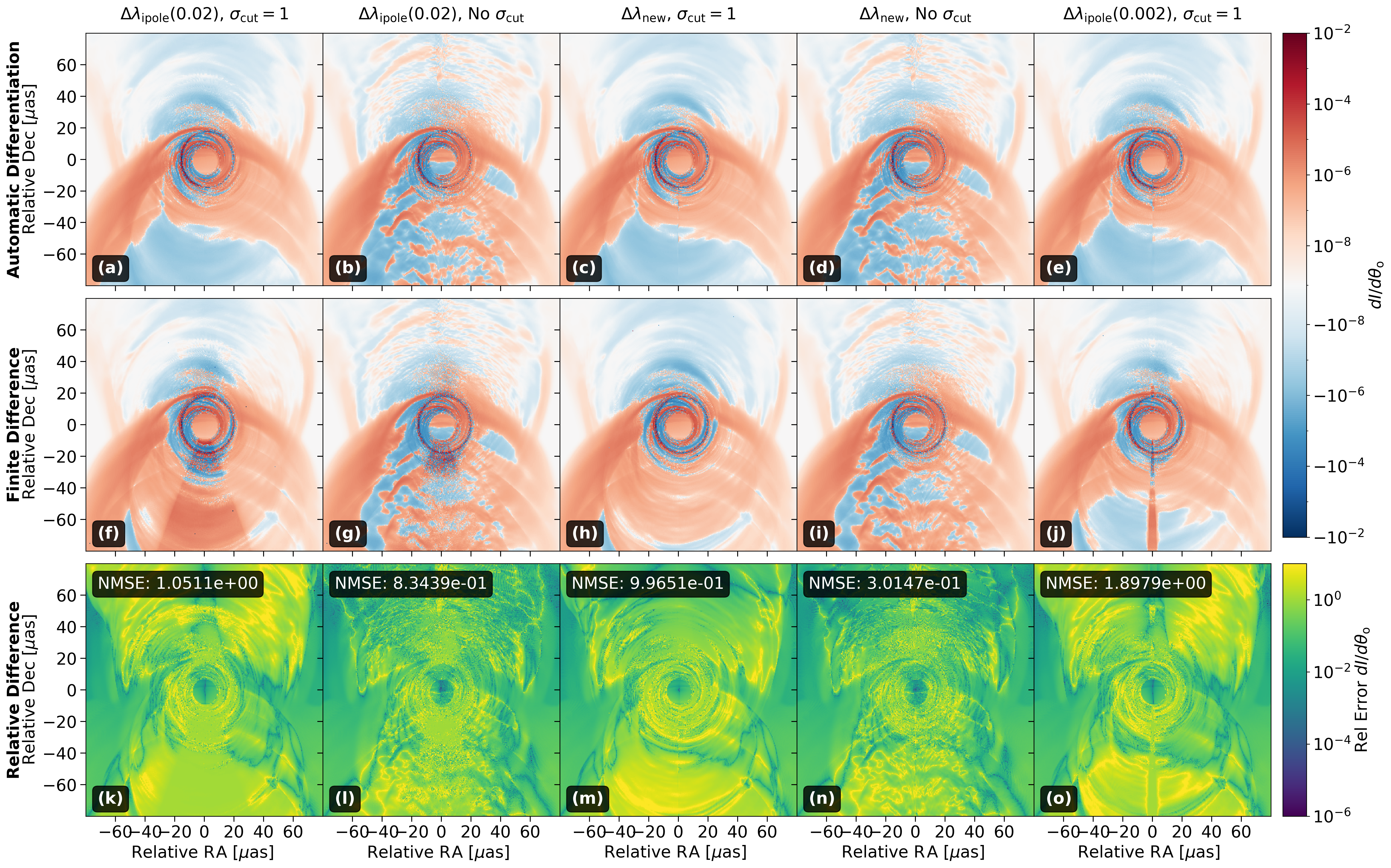}
    \caption{The sensitivity $dI/d\theta_o$ computed using different combinations of step size prescriptions and magnetization cutoffs. The top and middle rows correspond to the automatic differentiation (AD) and finite differences (FD) algorithms, respectively, while the bottom row shows the relative difference along with the NMSE. Each column corresponds to a different combination of step size prescription and magnetization cutoff, as indicated by the headers. The step size $\Delta \lambda_{\rm ipole}$ corresponds to the choice presented in Equations~\eqref{eq:stepsize_ipole}, where the value in parentheses denotes the poloidal angle cutoff ($\theta_{\rm cut}$) at which the step size begins to rapidly decrease near the pole, while $\Delta \lambda_{\rm new}$ corresponds to the choice of step size presented in Equations~\eqref{eq:stepsize_grmonty}. These images have a resolution of $320\times 320$.}
    \label{fig:different_stepsizes_sigmas}
\end{figure*}

Figure~\ref{fig:different_stepsizes_sigmas} shows the pixel-wise images for the observer angle sensitivity ($dI/d\theta_{\rm{o}}$) computed using AD and FD for different combinations of step size prescriptions, $\Delta \lambda$, and magnetization cutoffs, $\sigma_{\rm cut}$. The first two rows display the derivatives, while the third row shows the relative difference between them along with the corresponding NMSE.

We begin by noting the presence of a conical region in the south polar axis of the FD method when using $\Delta\lambda_{\rm ipole}$ in combination with $\sigma_{\rm cut} = 1$, as shown in Panel~\ref{fig:different_stepsizes_sigmas}(f). This region is a numerical artifact resulting from the simultaneous effect of the smaller step size and the $\sigma_{\rm cut}$ threshold. Importantly, this effect does not appear when $\Delta\lambda_{\rm ipole}$ and $\sigma_{\rm cut}$ are applied individually. 

Panel~\ref{fig:different_stepsizes_sigmas}(g) shows that using $\Delta\lambda_{\rm ipole}$ without $\sigma_{\rm cut}$ still produces a fuzzy, noisy cone funnel in regions where $\theta < \theta_{\rm cut}$. Turning to Panel~\ref{fig:different_stepsizes_sigmas}(h), we examine the effect of $\sigma_{\rm cut}$ when using $\Delta\lambda_{\rm new}$. While the conical region and fuzzy pattern associated with $\Delta\lambda_{\rm ipole}$ are absent, the results still differ from the AD calculations under the same conditions (Panel~\ref{fig:different_stepsizes_sigmas}(c)), particularly near the southern pole. When $\sigma_{\rm cut}$ is not applied, both FD and AD images agree well, with a normalized mean squared error of $\rm{NMSE} \sim 0.3$.

To test the effect of changing $\theta_{\rm cut}$, we reduce it by a factor of $10$, as shown in the fifth column of panels. This modification significantly changes the southern region of the FD images, making them more similar to the AD sensitivities. A narrower cone remains due to the step-size reduction, but it now appears in a thinner slab. The AD sensitivities remain consistent across Panels~\ref{fig:different_stepsizes_sigmas}(a,c,e), suggesting that automatic differentiation provides a more precise evaluation.

No clear numerical artifacts are visible in the derivative images in the third column of Figure~\ref{fig:different_stepsizes_sigmas}, unlike the configurations using $\Delta\lambda_{\rm ipole}$, where conical sharp regions or noisy funnel features appear near the poles. Nevertheless, large discrepancies between the AD and FD sensitivities remain, particularly in the southern hemisphere, where the FD solution shows deviations from the AD result.

Finally, we highlight that these numerical artifacts are not present in the calculation of $dI/dR_{\rm high}$, as these are artifacts of the geodesic integration. We also highlight that we have tested that these choices also work well for other GRMHD snapshots with different underlying resolution and simulation parameters. 

\section{The Landscape of The Model Parameters}
\label{sec:error_landscape}

In this section, we study what NMSE (Equation~\ref{eq:NMSE}) is generated when varying $\theta_{\rm{o}}$ and $R_{\rm high}$, i.e., the error landscape. This serves as a proxy for determining the difficulty of fitting the data, and does not require the gradient functionality of \code{Jipole}, which we will exploit in the next section. First, we examine the dependence of the error on $\theta_{o}$, while holding $R_{\rm high}$ fixed. Second, we analyze the error as a function of $R_{\rm high}$ for a fixed $\theta_{o}$. Finally, we present a two‑dimensional map of the error landscape when both $\theta_{o}$ and $R_{\rm high}$ vary simultaneously. 

\subsection{The Observer's Inclination}

We now generate images using the same GRMHD snapshot as in Section~\ref{sec:validation}, with $R_{\rm high} = 20$ for three observer angles: $\theta_{\rm{o}} = 60^\circ, 90^\circ, 163^\circ$. These images will serve as the ground truth and will be compared to images with observer angles ranging from $1^\circ$ to $179^\circ$, in increments of $1^\circ$. 

\begin{figure*}
    \centering
    \includegraphics[width=\textwidth]{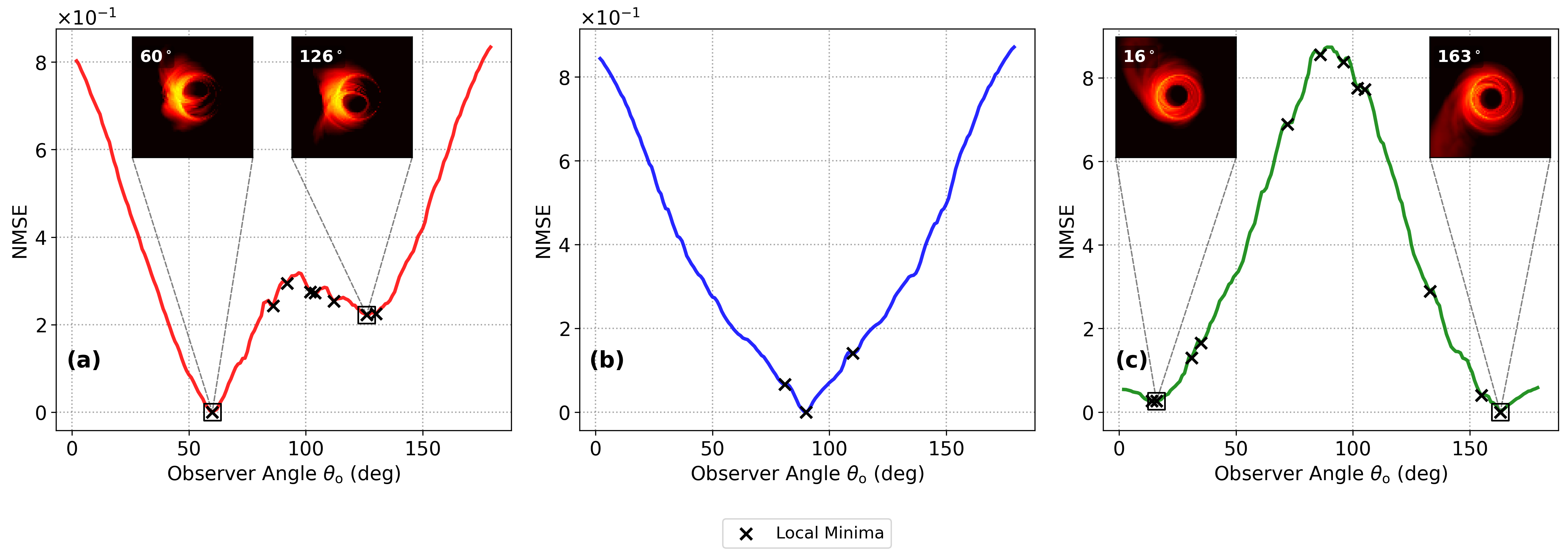}
    \caption{The normalized mean squared error (NMSE) between images at an angle $\theta_{\rm{o}}$ and an image at $60^\circ$, $90^\circ$, $163^\circ$ in panels (a), (b) and (c), respectively, all with $R_{\rm high} = 20$. The presence of a local minimum is denoted as a black cross in every panel. For panels (a) and (c), we show the corresponding image (in logarithmic scale) at the highlighted local minimum (i.e., $60^\circ$ and $126^\circ$ for panel (a), and $16^\circ$ and $163^\circ$ for panel (c)).}
    \label{fig:Thetao_error}
\end{figure*}

As shown in Figure~\ref{fig:Thetao_error}, for a given reference observer angle, the error landscape typically exhibits two minima. One minimum occurs at the reference angle itself, corresponding to the true observer view, while a secondary local minimum appears near the supplementary angle $180^\circ - \theta_{\rm{o}}$ due to the poloidal similar symmetry of the geodesics, as evidenced by panels (a,c). The gas distribution above and below the disk is similar, meaning the NMSE does not change drastically between these symmetric views. However, the minima are not located exactly at $180^\circ - \theta_{\rm{o}}$; they are slightly shifted because small differences in the gas along the line of sight and in the geodesics break perfect symmetry, leading to a deviation of the local minimum from the exact mirrored angle.

\subsection{The Electron Heating Model Parameter}

We now focus on how changing the electron heating parameter $R_{\rm high}$ affects the NMSE for different reference values. As expected, given that the photon trajectories are not modified, this error landscape does not exhibit multiple minima with respect to $R_{\rm high}$, as shown in Figure~\ref{fig:Thetao_error}. However, there are significant differences in the gradient of the error function between low and high $R_{\rm high}$ values. At low $R_{\rm high}$, the NMSE changes rapidly with small variations in the parameter, whereas at high $R_{\rm high}$, the error landscape flattens considerably. This disparity in gradient magnitude can pose challenges when adapting the step size in a search algorithm: steps that are appropriate for the steep low-$R_{\rm high}$ region may be too aggressive for the flat high-$R_{\rm high}$ region, potentially slowing convergence or causing instability when finding the best-fit. 

\begin{figure*}
    \centering
    \includegraphics[width=\textwidth]{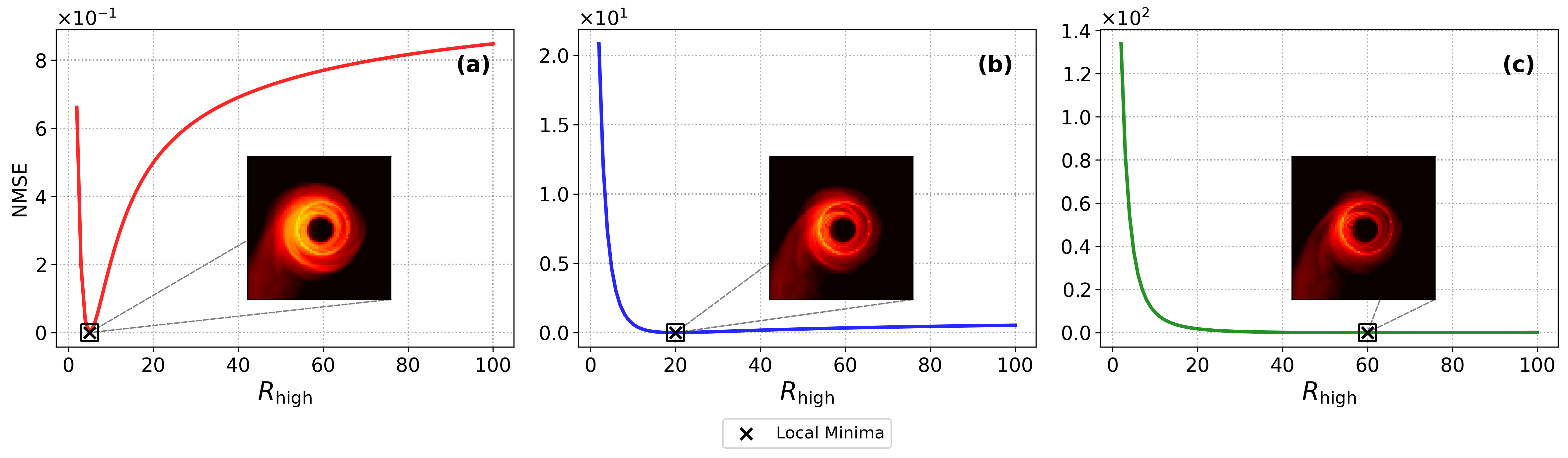}
    \caption{The normalized mean squared error (NMSE) between images at a given $R_{\rm high}$ and the ground truth image at $R_{\rm high} = 5$, $20$, and $60$ in panels (a), (b), and (c), respectively, all with $\theta_{\rm{o}} = 163^\circ$. The presence of a local minimum is denoted as a black cross in every panel. On each panel we show the corresponding image (in logarithmic scale) for the global minimum of each $R_{\rm high}$ curve ($5$, $20$, and $60$) in its respective panel.}
    \label{fig:Rhigh_error}
\end{figure*}

\subsection{Joint Error Landscape}

\begin{figure}
    \centering
    \includegraphics[width=\linewidth]{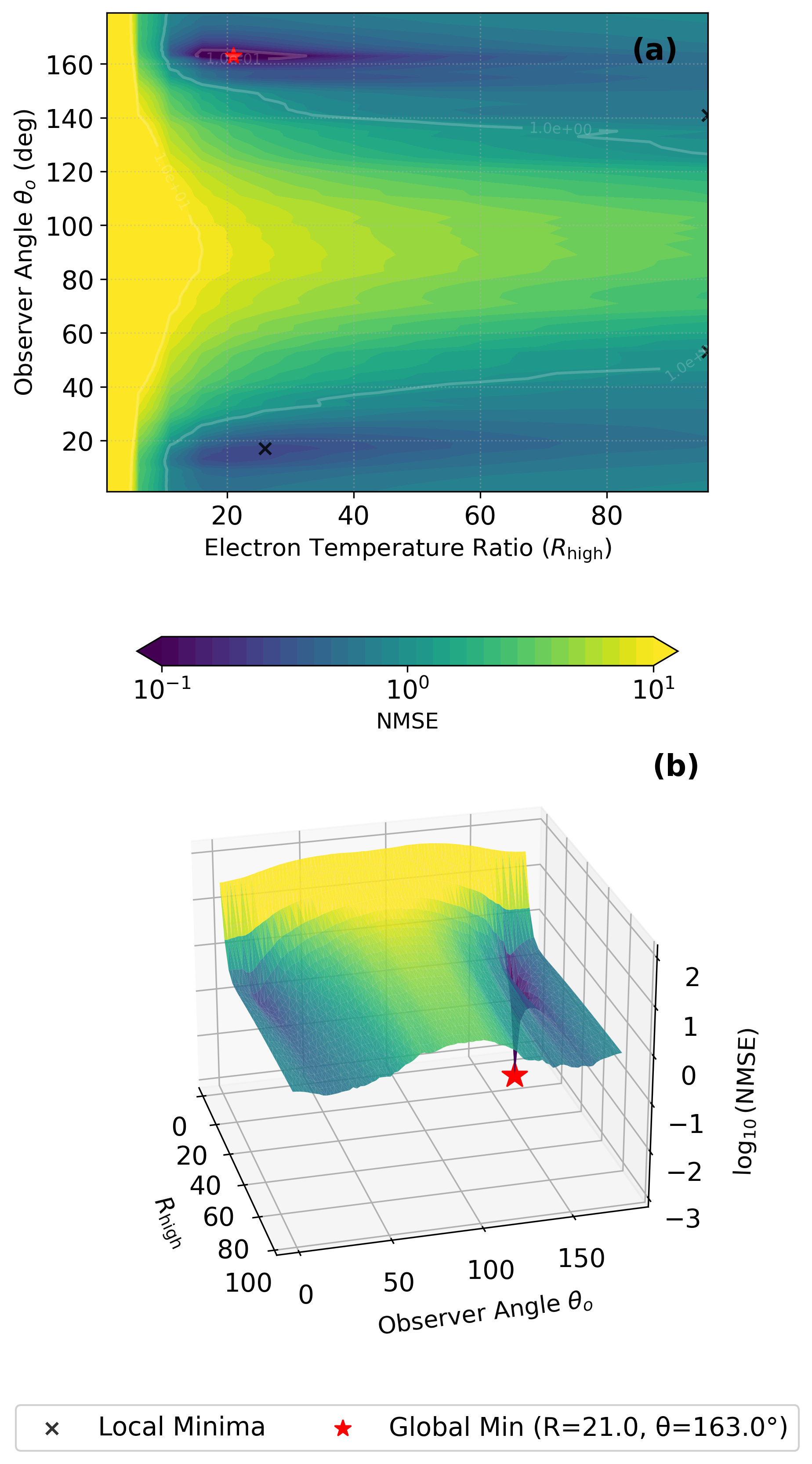}
    \caption{The behavior (landscape) of the NMSE across the parameter space of electron temperature ratio ($R_{\rm high}$) and observer angle ($\theta_{\rm{o}}$). (a) A 2D logarithmic contour map illustrating the error distribution. The global minimum is indicated by the red star at $R_{\rm high} = 21.0$ and $\theta_{\rm{o}}= 163.0^\circ$, while secondary local minima are marked with black crosses. (b) A $3$D surface plot of the same parameter space showing the $\log_{10}$-transformed NMSE. Both panels utilize a logarithmic color mapping to emphasize error variations across orders of magnitude.}
    \label{fig:grid_error_3d}
\end{figure}

Let us now consider the simultaneous variation of the observer inclination $\theta_{\rm{o}}$ and the electron temperature ratio $R_{\mathrm{high}}$, considering the ground truth values of $\theta_{\rm{o}} = 163^\circ$ and $R_{\rm high} = 20$. This more complex scenario is closer to an actual data analysis case, where several parameters must be studied simultaneously. For this combined case, we use a step size of $2^\circ$ for $\theta_{\rm{o}}$ and five for $R_{\rm high}$. Figure~\ref{fig:grid_error_3d} illustrates the NMSE landscape as a $2$D surface and a $3$D log-scaled projection. As shown in these figures, their coupled effect on the image morphology can lead to degeneracies that complicate the optimization process.

Consistent with the behavior detailed in the previous section, lower values of $R_{\rm high}$ have a substantial impact on the overall error. We also observe a distinct peak in the error landscape between the ground truth and the region near its supplementary angle. Conversely, increasing $R_{\rm high}$ considerably reduces the error across the intermediate viewing angles, specifically within the range of $\theta_{\rm{o}} \approx 20^\circ - 140^\circ$.

Furthermore, the parameter space shown here exhibits a noticeably lower count of local minima. This is a direct consequence of the larger step sizes employed for this specific grid evaluation, which were set to $2^\circ$ for $\theta_{\rm{o}}$ and $5$ for $R_{\rm high}$. Because the grid for the electron temperature ratio starts at $1$ and increases in increments of $5$, the apparent global minimum for $R_{\rm high}$ is recorded as $21$ in the figure, representing the closest sampled grid point to the true optimal value of $R_{\rm high} = 20$.

\section{Simple Mock Data Analyses Examples}
\label{sec:MockData}

Having established the correctness of \code{Jipole} with GRMHD simulations and understood how the studied model parameters modify the resulting images, we will now use these sensitivities to inform parameter estimation. Thus, following closely \cite{NaetheMotta_2025_jipole}, we will use these sensitivities in a conjugate gradient (CG) optimization method as a proof of concept rather than a finalized fitting framework for black hole imaging data analysis. In particular, the current implementation does not account for systematic uncertainties, which are essential when dealing with realistic inference. We leave that analysis for future work. 

\subsection{The Conjugate Gradient Method}
\label{sec:CG}

We follow the CG optimization framework introduced in \citep{NaetheMotta_2025_jipole}, but with few modifications to deal with the resulting error landscape obtained from imaging GRMHD simulations. In brief, we seek to minimize the NMSE between the reference image and the model image by iteratively updating the model parameters along descent directions informed by the sensitivities. At each iteration, the CG method constructs a search direction that combines the local gradient of the cost function with information from previous steps, and a backtracking Armijo line search with bound constraints is used to determine an appropriate step size along this direction that sufficiently reduces the cost function. To improve numerical conditioning, the optimization is performed in a scaled parameter space, where $\theta_o$ and $R_{\rm high}$ are normalized by factors of $100$ and $200$, respectively. The modifications described below focus on improving the previous algorithm.

The primary modification to the CG scheme employed in \citep{NaetheMotta_2025_jipole} involved dynamically adjusting the CG step size based on the results of the previous line search. At the first iteration, the step size is initialized such that parameter updates do not exceed $\Delta \theta_{\rm max} = 20^\circ$ and $\Delta R_{\rm high, max} = 10$. Lower bounds are imposed to prevent excessively small updates, ensuring minimum variations of $5^\circ$ in $\theta_o$ and $5$ in $R_{\rm high}$. When a successful step is found, the step size for the next iteration is adjusted based on how much the search direction has changed. In practice, the previous step size is scaled by the ratio of the old and new directional derivatives. This scaling factor is clamped to a fixed range, between $0.1$ and $5.0$, to prevent excessively large or small updates. 

In cases where the line search fails to identify a suitable step, the initialization procedure is reset using more conservative limits, with maximum allowed variations reduced to $\Delta \theta_{o, {\rm max}} = 10^\circ$ and $\Delta R_{\rm high, max} = 10$.

In contrast to the analytical emission model used in~\cite{NaetheMotta_2025_jipole}, an additional basin-hopping–inspired strategy needs to be introduced to improve the optimization method. As we will show below, the error landscape exhibits some local minima, in which a standard CG method may become trapped due to its inherently local, descent-based nature. This behavior can stop convergence to the global minimum and was not encountered in the simpler analytical setting considered previously.

To tackle this issue, a stagnation-detection mechanism is implemented: when the improvement in the cost function falls below a prescribed threshold, the algorithm probes the parameter space in a set of discrete directions around the current point. Trial steps are taken independently along each parameter axis, with initial step sizes of $\Delta \theta_o = 5.0^\circ$ and $\Delta R_{\rm high} = 5.0$, while enforcing the prescribed parameter bounds. The corresponding cost function values are then evaluated for each trial direction. If a lower-cost configuration is identified, the optimization is reinitialized from this new point, allowing the algorithm to escape shallow local minima. If no improvement is found, the probing step sizes are reduced by a factor of $0.8$, and the procedure is repeated for up to $15$ rounds. If no improvement is achieved after all probing attempts, the algorithm is terminated.

We emphasize that this algorithm could be enhanced with additional features to determine optimal step sizes, such as logarithmic parameterization or advanced preconditioning. We could also choose a more effective scaling between parameters to accommodate varying gradient scales. However, since our primary objective is to demonstrate the application of these sensitivities, there is no necessity to improve the implemented fitting algorithm.

With the optimized tool defined, we will now perform fits using images with a resolution of $128 \times 128$ under different conditions. 

\subsection{Noise-Free and Blur-Free Parameter Fitting}

To start our analysis, we first generate a ground-truth image using $\theta_{\rm{o}} = 163^\circ$ and $R_{\rm high} = 20$. We then perform a sequence of fits: first estimating $\theta_{\rm{o}}$ alone, followed by $R_{\rm high}$, and finally carrying out a joint fit of both parameters.

\begin{figure}
    \centering
    \includegraphics[width=\linewidth]{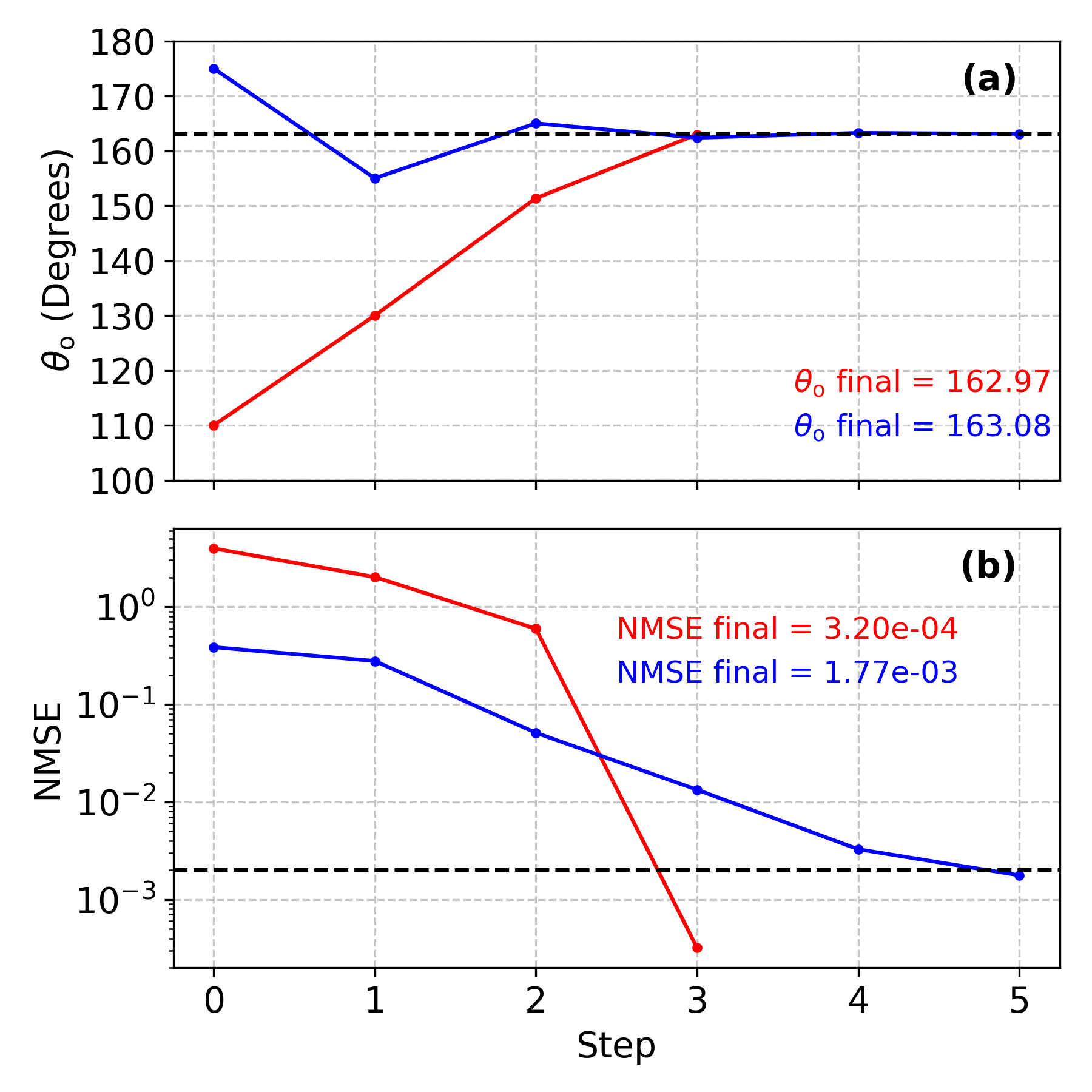}
    \caption{Panel (a): The evolution of the observer's inclination, $\theta_{\rm{o}}$, as a function of the iteration step during the CG optimization. The black dashed line corresponds the true reference value. The blue line corresponds to an initial guess of $175^\circ$, while the red line corresponds to a guess of $110^\circ$. Panel (b): The evolution of the NMSE per step. The black dashed line corresponds to the chosen tolerance value to stop optimizing.}
    \label{fig:th_clean_CG}
\end{figure}

The fitting of the $\theta_{\rm{o}}$ parameter is illustrated in Figure~\ref{fig:th_clean_CG}. We observe rapid convergence in both cases, whether the initial value is chosen above or below the reference value, reaching the tolerance cost of $2 \times10^{-3}$ in fewer than $10$ steps. We intentionally choose initial values to the right of the peak shown in Figure~\ref{fig:Thetao_error}(c). If values on the left side of the peak are used instead, the fit tends to converge to the local minimum at approximately $\theta_{\rm{o}} \sim 16$. We acknowledge that this behavior arises from the chosen optimization method and emphasize that this simple approach won't be generically employed when fitting real observational data. 

\begin{figure}
    \centering
    \includegraphics[width=\linewidth]{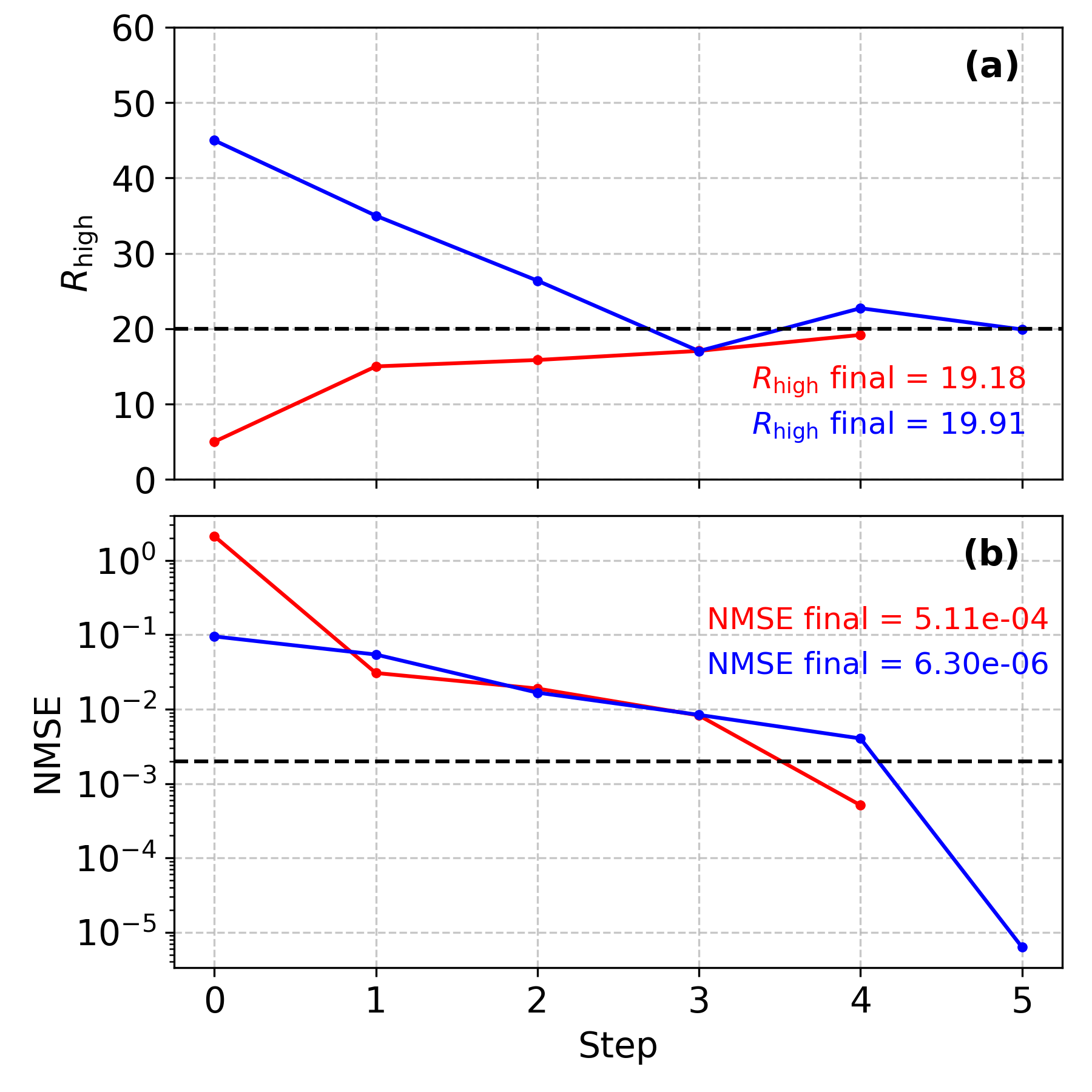}
    \caption{The evolution of $R_{\mathrm{high}}$ as a function of the iteration step during the CG optimization. The blue line corresponds to an initial guess of $45$, the red line corresponds to a guess of $5$, and the black dashed line shows the true reference value. Panel (b): The evolution of the NMSE as a function of iteration. The black dashed line indicates the tolerance value used to stop the optimization.}
    \label{fig:Rhigh_clean_CG}
\end{figure}

The next step is to visualize the fitting of the $R_{\rm high}$ parameter, illustrated in Figure~\ref{fig:Rhigh_clean_CG}. We start the fits from initial guesses of $5$ and $45$. Similarly to the $\theta_{\rm{o}}$ case, we observe rapid convergence toward the target solution, with the cost function reaching the tolerance threshold of $2 \,\times\, 10^{-3}$ in fewer than ten steps for both higher and lower initial guesses.

\begin{figure}
    \centering
    \includegraphics[width=\linewidth]{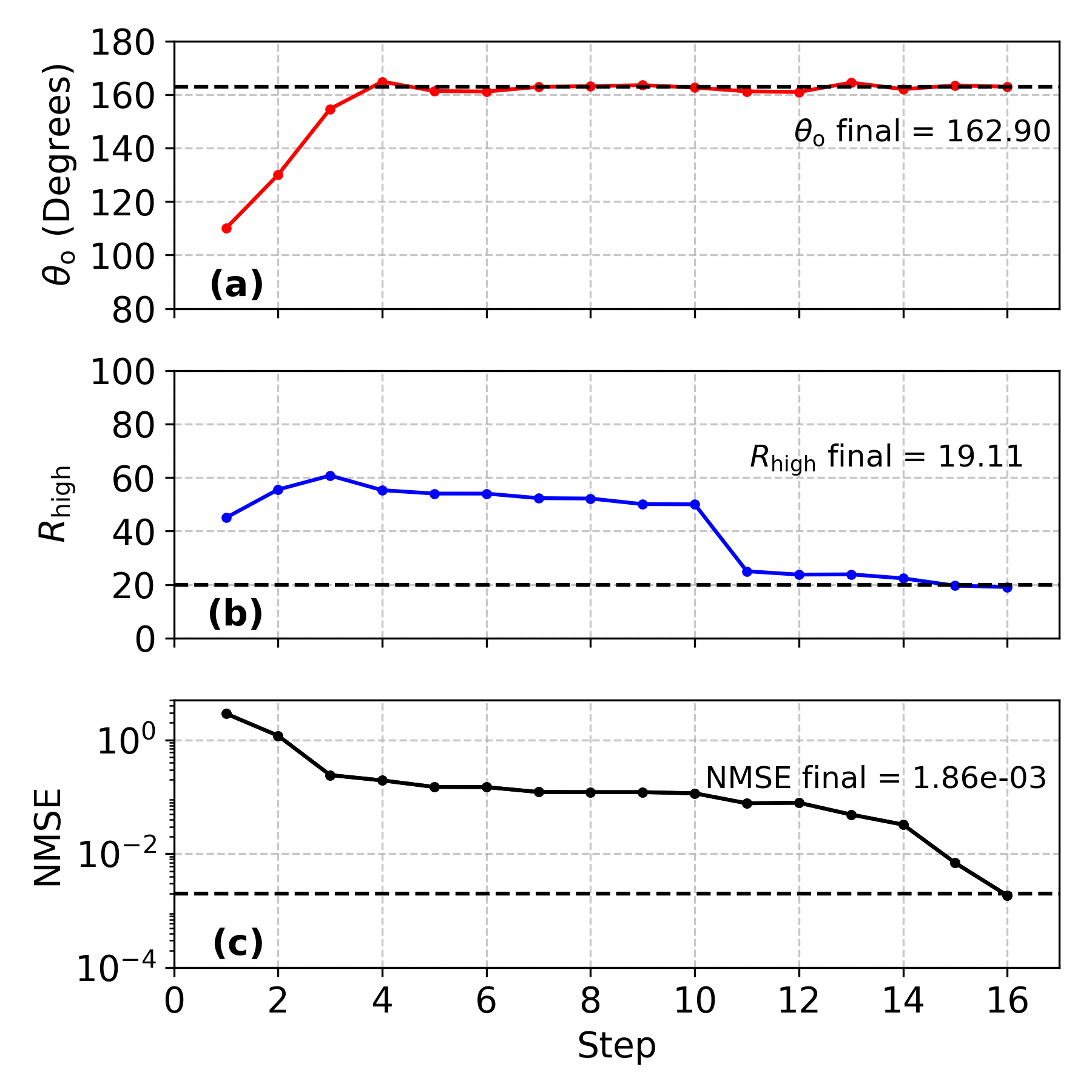}
    \caption{The evolution of $\theta_{\rm{o}}$ and $R_{\rm high}$ using the CG optimization when the two parameters are evolved jointly. 
    Panel (a) shows the evolution of $\theta_{\rm{o}}$ per iteration step, with the black dashed line indicating the true reference value. 
    Panel (b) shows the evolution of $R_{\rm high}$ per iteration, with the black dashed line indicating the true reference value. 
    Panel (c) displays the NMSE of the image per iteration, with the black dashed line representing the tolerance used to stop the optimization.}
    \label{fig:both_clean_CG}
\end{figure}

We now perform a joint fit of both parameters, exploring the error landscape shown in Figure~\ref{fig:grid_error_3d}, and present the optimization results in Figure~\ref{fig:both_clean_CG}. Both $\theta_{\rm{o}}$ and $R_{\rm high}$ converge to values close to their true reference points, indicated by the dashed black lines. The NMSE decreases steadily throughout the iterations, ultimately falling below the stopping tolerance, which demonstrates that the joint fitting procedure successfully recovers the target parameters with accuracy. In particular, the fitted value of $R_{\rm high}$, expected to be $20$, converges to approximately $19.11$, while $\theta_{\rm{o}}$ approaches its true value closely, reaching $162.9^\circ$. 

We also observe an increase in the fitted value of $R_{\rm high}$ when $\theta_{\rm{o}}$ lies approximately in the range $110^\circ \lesssim \theta_{\rm{o}} \lesssim 150^\circ$. As evidenced by Figure~\ref{fig:grid_error_3d}, in this region of parameter space an increase in $R_{\rm high}$ leads to a decrease in the error, indicating a local degeneracy between the two parameters.

The joint optimization requires a larger number of iterations compared to the single-parameter fits. This is primarily because $R_{\rm high}$ initially explores higher values and decreases only gradually toward its optimal value near $20$. This behavior reflects differences in the gradients of the error with respect to each parameter: the error surface is relatively shallow along the $R_{\rm high}$ direction (Figure~\ref{fig:Rhigh_error}), leading to smaller updates and slower convergence, whereas the gradient with respect to $\theta_{\rm{o}}$ is steeper (Figure~\ref{fig:Thetao_error}), allowing it to converge more rapidly.

\subsection{Parameter Fitting in the Presence of Gaussian Noise and Image Blurring}
\label{sec:blur_noise_fitting}

We now consider a more realistic (though still highly idealized) scenario by incorporating both blurring and noise into the synthetic image. We keep the fitting procedure as in the previous sections. 

As a proxy for instrument resolution, the modeled images are convolved with a Gaussian kernel $G_{\sigma}$, defined as
\begin{equation}
    I_{\rm blur} = I_{\rm truth} * G(\sigma),
    \label{Eq:gaussian_blur}
\end{equation}
where $\sigma$ is the standard
deviation parameter
\begin{equation}
    \sigma = \frac{\frac{N}{2\theta_{\mu\rm{as}}} \cdot 20}{2\sqrt{2\ln 2}}.
\end{equation}
Here, the symbol $*$ denotes convolution, $\theta_{\mu\rm{as}}$ is the field of view in microarcseconds, and $N$ is the number of pixels along one dimension of the image array. The blurred image is then transformed into the Fourier domain, $V = \mathcal{F}[I_{\mathrm{blur}}]$, where $\mathcal{F}$ denotes the two-dimensional Fourier transform. Working in the visibility domain is particularly appropriate for interferometric observations, since VLBI measurements correspond to complex visibilities rather than direct image intensities.

To set the noise level, we compute a characteristic scale for the visibilities using their root-mean-square value,
\begin{equation}
V_{\mathrm{rms}} = \mathrm{std}(V),
\end{equation}
where $\mathrm{std}(\cdot)$ represents the standard deviation. We adopt a signal-to-noise ratio of $\mathrm{SNR} = 15$, which defines the noise amplitude as
\begin{equation}
\sigma_{\mathrm{noise}} = \frac{V_{\mathrm{rms}}}{\mathrm{SNR}}.
\end{equation}

Complex Gaussian noise is then generated with zero mean and variance $\sigma_{\mathrm{noise}}^2$ in both the real and imaginary components, and added directly to the visibilities, yielding
\begin{equation}
V_{\mathrm{noisy}} = V + N.
\end{equation}

Finally, the noisy visibilities are transformed back into the image domain via the inverse Fourier transform,
\begin{equation}
I_{\mathrm{noisy}} = \mathrm{Re}\left[\mathcal{F}^{-1}(V_{\mathrm{noisy}})\right],
\end{equation}
where $\mathrm{Re}(\cdot)$ denotes the real part. This reconstructed image is then used as input for the fitting procedure, where the conjugate gradient method is employed to recover the model parameters.

\begin{figure}
    \centering
    \includegraphics[width=\linewidth]{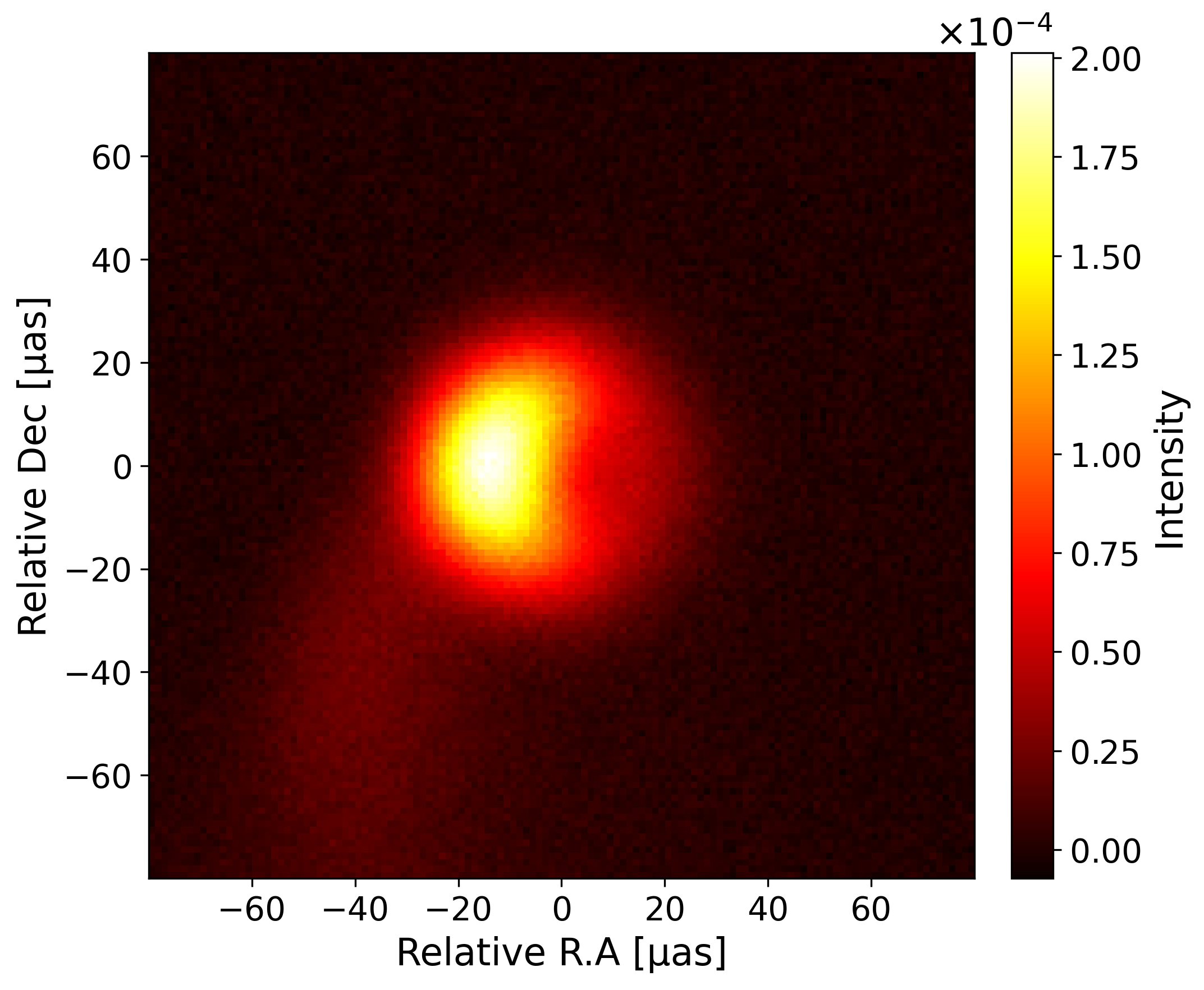}
    \caption{The resulting image after the inclusion of blurring and noise, as described in Section~\ref{sec:blur_noise_fitting}. This corresponds to Figure~\ref{fig:GRMHD_comparison} after convolution with a Gaussian beam of $20\,\mu\mathrm{as}$ full width at half maximum, and the addition of a complex Gaussian noise with a signal-to-noise ratio of $15$ in the visibility domain before transforming back to the image domain.}
    \label{fig:I_blur_noise}
\end{figure}

The resulting image with blurring and noise is shown in Figure~\ref{fig:I_blur_noise}. It is important to note that throughout the optimization, the model image is convolved with the Gaussian kernel at each iteration to consistently incorporate the effects of finite instrumental resolution. The noise, however, is not reintroduced during the fitting steps; it is added only once to the synthetic data that serve as the reference.

\begin{figure}
    \centering
    \includegraphics[width=\linewidth]{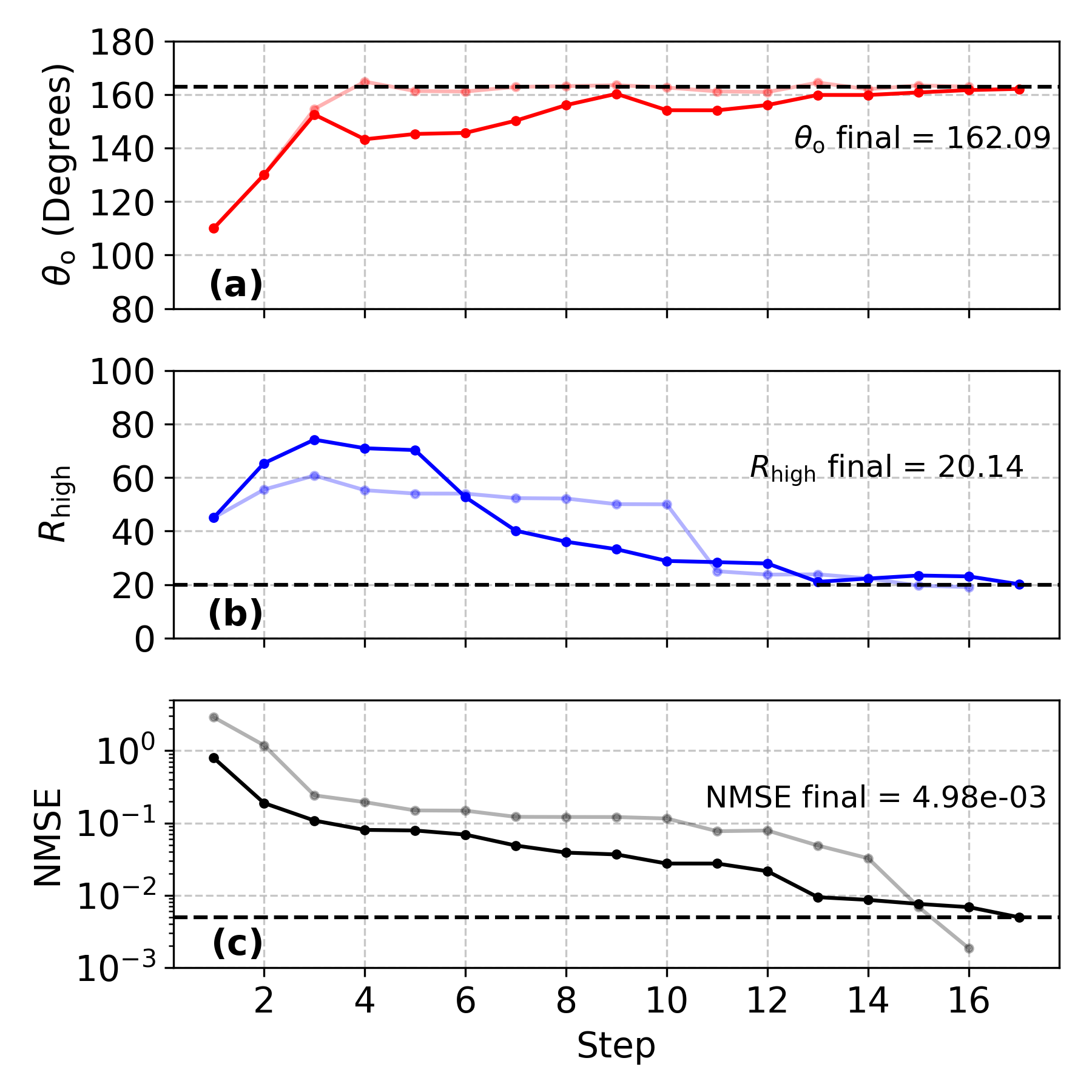}
    \caption{Joint fitting of $\theta_{\rm{o}}$ and $R_{\rm high}$ under noise and blur using the CG optimization. The transparent lines refer to the fitting without blur and noise portrayed in Figure~\ref{fig:both_clean_CG}. 
    Panel (a) shows the evolution of $\theta_{\rm{o}}$ per iteration step, with the black dashed line indicating the true reference value. 
    Panel (b) shows the evolution of $R_{\rm high}$ per iteration, with the black dashed line indicating the true reference value. 
    Panel (c) displays the NMSE of the image per iteration, with the black dashed line representing the tolerance used to stop the optimization.}
    \label{fig:both_blurnoise_CG}
\end{figure}

As an analysis of how the method would evolve in an environment with blur and noise, we again perform a joint fit of $R_{\rm high}$ and $\theta_{\rm{o}}$ in Figure~\ref{fig:both_blurnoise_CG}. For this fit, we increase the error tolerance to $5 \,\times\, 10^{-3}$, since the blur and noise will affect the NMSE found even at the correct ground truth values. We observe a relatively fast convergence within $20$ iteration steps. 

\section{Discussion}
\label{sec:Discussion}

In this work, we have used \code{Jipole} to compute, for the first time, image sensitivities from general relativistic magnetohydrodynamics (GRMHD). We have established that these sensitivities can be computed in a stable and physically meaningful way in a realistic simulation setting, and showed the information they provide about image formation. As a concrete demonstration of their use, we employed them to perform simple fitting of image parameters given a simulated reference image. We considered two types of parameters: the observer’s inclination $\theta_{\rm{o}}$, a geometrical parameter, and the electron heating model parameter $R_{\rm high}$, an astrophysical parameter. These two parameter types are representative of all imaging parameters, and the method is easily extensible to include any additional parameters. 

We have also validated the generation of thermal synchrotron images of GRMHD snapshots in \code{Jipole} against \code{ipole}. Using the same $3$D SANE simulation presented in \cite{Prather_2023}, we obtain agreement between both codes, with a normalized mean square error (NMSE) of $\sim 10^{-13}$. We have further compared the sensitivities $dI/d\theta_{\rm{o}}$ and $dI/dR_{\rm high}$ between our automatic differentiation implementation and finite difference calculations. Based on these comparisons, we detail the technical implications of utilizing different integration step sizes and magnetization cutoffs, revealing the best agreement between the FD and AD when no magnetization cutoff is applied.

To better understand the feasibility of parameter searches guided by image sensitivities, we mapped the error landscape of our model parameters. By evaluating the NMSE as a function of $\theta_{\rm{o}}$ and $R_{\rm high}$, both independently and jointly, we showcase some topological features of the GRMHD image error surface. Firstly, we observe a symmetry-induced local minimum in $\theta_{\rm{o}}$ at the supplementary angle ($ 180^\circ - \theta_{\rm{o}}$). This is driven by the approximate poloidal symmetry of the gas distribution, creating an error peak that can easily trap descent-based optimization algorithms. Secondly, the landscape showcases asymmetric gradient magnitudes in $R_{\rm high}$.  The NMSE changes drastically at low values of $R_{\rm high}$ but flattens out considerably at higher values, complicating step-size selection for iterative solvers. This behavior highlights that the low-$R_{\rm high}$ regime is both highly sensitive and structurally complex, making it a critical region for parameter inference and one that benefits strongly from gradient-informed exploration. Finally,  when mapped simultaneously, the joint influence of $\theta_{\rm{o}}$ and $R_{\rm high}$ introduces some degeneracies. Lower $R_{\rm high}$ values dominate the overall error, while higher values flatten the error across intermediate viewing angles.

To demonstrate the practical utility of these sensitivities, a conjugate gradient algorithm was employed to recover parameters from a ground truth image generated with $\theta_{\rm{o}} = 163^\circ$ and $R_{\rm high} = 20$. The fitting procedure was carried out systematically: first by optimizing $\theta_{\rm{o}}$ in isolation, then focusing solely on $R_{\rm high}$, and concluding with a joint fit of both parameters. In all three test cases, the algorithm successfully converged to the exact target values within a small number of iterations. To further evaluate the robustness of this gradient-based approach under realistic observational conditions, the optimizations were repeated under noise and blur, matching the resolution constraints and signal-to-noise ratios typical of actual interferometric data. In this scenario, we observe that the method still successfully fits the ground truth parameters.

In theory, and especially since the present implementation has not yet been optimized, the computation of derivatives via automatic differentiation is significantly more expensive than generating a single image. However, the growth in computational complexity when evolving differentials with respect to additional parameters is slower than the computation required to run one additional image per parameter, as in finite-difference methods. As the parameter space expands, inline computation of derivatives should become increasingly advantageous, especially if significant optimization of the current codebase is possible, as we expect. Future work will focus on optimizing the AD pipeline, particularly in preparation for deployment on clusters, where we expect substantial reductions in wall-clock time.

As we have mentioned, the conjugate gradient method adopted here is intended only as an illustrative test of whether AD-computed differential images can support parameter recovery and reduce the reliance on large image libraries. Our next step is to integrate \code{Jipole} into full sampling frameworks such as \code{Comrade.jl} \citep{Tiede_2022}, enabling gradient-informed inference under realistic observational conditions. Such an approach would allow model-data comparisons to be guided directly in parameter space and could improve the efficiency of analyses of current and future VLBI observations \citep{Johnson:2023ynn,Johnson:2024ttr}.

\begin{acknowledgments}

We thank Charles Gammie, Daniel Palumbo, Hyun Lim, Rodrigo Nemmen and Maur\'icio Richartz for helpful discussions. P. N. M. was supported by FAPESP (Fundação de Amparo à Pesquisa do Estado de São Paulo) under grant 2023/15835-2. M. R. N. acknowledges UFABC, UNIVESP and UNICID for the support. 
C.P. was supported by the Black Hole Initiative at Harvard University, which is funded in part by the Gordon and Betty Moore Foundation (Grant \#13526). It was also made possible through the support of a grant from the John Templeton Foundation (Grant \#63445). The opinions expressed in this publication are those of the author(s) and do not necessarily reflect the views of these Foundations.
This work used Delta at NCSA through allocation PHY250091 from the Advanced Cyberinfrastructure Coordination Ecosystem: Services $\text{\&}$ Support (ACCESS) program, which is supported by U.S. National Science Foundation grants $\text{\#2138259}$, $\text{\#2138286}$, $\text{\#2138307}$, $\text{\#2137603}$, and $\text{\#2138296}$.

\end{acknowledgments}

\bibliography{citations}{}
\bibliographystyle{aasjournal}

\end{document}